\documentclass[aps,prd,preprint,tightenlines,nofootinbib]{revtex4}
\usepackage{epsfig}
\usepackage{graphicx}
\usepackage{graphics}
\usepackage{dcolumn}
\usepackage{bm}
\usepackage{afterpage}
\usepackage{citesort}

\newcommand{\etal}{{\it et al.}}

\begin{document}

\preprint{\tighten\vbox{\hbox{\hfil EFI 11-34}}}
\title
{\LARGE Leptonic Decays of Charged Pseudoscalar Mesons -- 2012}

\author{Jonathan L. Rosner}
\affiliation{Enrico Fermi Institute, University of Chicago, Chicago, IL 60637}
\author{and Sheldon Stone}
\affiliation{Department of Physics, Syracuse University, Syracuse, NY 13244\\
\\}
\date{\today}

\begin{abstract}
We review the physics of purely leptonic decays of $\pi^\pm$, $K^\pm$,
$D^{\pm}$, $D_s^\pm$, and $B^\pm$ pseudoscalar mesons.  The measured decay
rates are related to the product of the relevant weak-interaction-based CKM
matrix element of the constituent quarks and a strong interaction parameter
related to the overlap of the quark and antiquark wave-functions in the meson,
called the decay constant $f_P$. The interplay between theory and experiment is
different for each particle. Theoretical predictions of $f_B$ that are needed
in the $B$ sector can be tested by measuring $f_{D^+}$ and $f_{D_s^+}$ in the
charm sector.  The lighter $\pi^{\pm}$ and $K^{\pm}$ mesons provide stringent
comparisons between experiment and theory due to the accuracy of both the
measurements and the theoretical predictions.  This review was prepared for the
Particle Data Group's 2012 edition \cite{Previous}.
\end{abstract}
\maketitle

\section{Introduction}
\begin{figure}[hbt]
\centering
\includegraphics[width=3in]{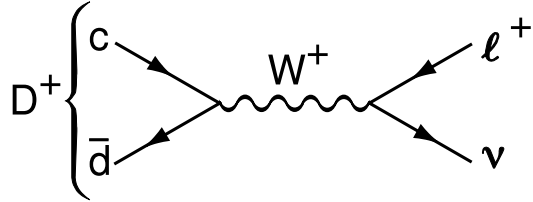}\vskip -0.02mm
\caption{The annihilation process for pure $D^+$ leptonic decays in the
Standard Model.
 } \label{Ptoellnu}
\end{figure}

Charged mesons formed from a quark and antiquark can decay to a
charged lepton pair when these objects annihilate via a virtual
$W$ boson. Fig.~\ref{Ptoellnu} illustrates this process for the
purely leptonic decay of a $D^+$ meson.

Similar quark-antiquark annihilations via a virtual $W^+$ to
the $\ell^+ {\nu}$ final states occur for the $\pi^+$, $K^+$, $D_s^+$, and
$B^+$ mesons.  (Charge-conjugate particles and decays are implied.) Let $P$ be
any of these pseudoscalar mesons.  To lowest order, the decay width is
\begin{equation}
\Gamma(P\to \ell\nu) = {{G_F^2}\over 8\pi}f_{P}^2\ m_{\ell}^2M_{P}
\left(1-{m_{\ell}^2\over M_{P}^2}\right)^2 \left|V_{q_1
q_2}\right|^2~. \label{equ_rate}
\end{equation}

\noindent Here $M_{P}$ is the $P$ mass, $m_{\ell}$ is the $\ell$
mass, $V_{q_1 q_2}$ is the Cabibbo-Kobayashi-Maskawa (CKM) matrix
element between the constituent quarks $q_1 \bar q_2$ in $P$, and
$G_F$ is the Fermi coupling constant. The parameter $f_P$ is the
decay constant, and is related to the wave-function overlap of the
quark and antiquark.

The decay $P^\pm$ starts with a spin-0 meson, and ends up with a
left-handed neutrino or right-handed antineutrino.  By angular
momentum conservation, the $\ell^\pm$ must then also be left-handed
or right-handed, respectively. In the $m_\ell = 0$ limit, the decay
is forbidden, and can only occur as a result of the finite $\ell$
mass.  This helicity suppression is the origin of the $m_\ell^2$
dependence of the decay width.

There is a complication in measuring purely leptonic decay rates.
The process $P\to \ell\nu\gamma$ is not simply a radiative
correction, although radiative corrections contribute. The $P$ can
make a transition to a virtual $P^*$, emitting a real photon, and
the $P^*$ decays into $\ell\nu$, avoiding helicity suppression. The
importance of this amplitude depends on the decaying particle and
the detection technique.  The $\ell\nu\gamma$ rate for a heavy
particle such as $B$ decaying into a light particle such as a muon
can be larger than the width without photon emission \cite{Bradcor}.
On the other hand, for decays into a $\tau^{\pm}$, the helicity
suppression is mostly broken and these effects appear to be small.

Measurements of purely leptonic decay branching fractions and
lifetimes allow an experimental determination of the product
$\left|V_{q_1 q_2}\right| f_{P}$. If the CKM element is well known
from other measurements, then $f_P$ can be well measured. If, on the
other hand, the CKM element is not well measured, having
theoretical input on $f_P$ can allow a determination of the CKM
element.   The importance of measuring $\Gamma(P\to \ell\nu)$
depends on the particle being considered. In the case of the $B^-$ the 
measurement of $\Gamma(B^-\to\tau^-\overline{\nu})$ provides an indirect 
determination of $|V_{ub}|$ provided that $f_B$ is provided by theory.
In addition, $f_B$
is crucial for using measurements of $B^0$-$\overline{B}^0$ mixing
to extract information on the fundamental CKM parameters.  Knowledge
of $f_{B_s}$ is also needed, but it cannot be directly
measured as the $B_s$ is neutral, so the violation of the SU(3)
relation $f_{B_s} = f_B$ must be estimated theoretically. This
difficulty does not occur for $D$ mesons as both the $D^+$ and
$D_s^+$ are charged, allowing the direct measurement of SU(3)
breaking and a direct comparison with theory.

For $B^-$ and $D_s^+$ decays, the existence of a charged Higgs boson
(or any other charged object beyond the Standard Model) would modify
the decay rates; however, this would not necessarily be true for
the $D^+$ \cite{Hou,Akeroyd}.  More generally, the ratio of $\tau \nu$
to $\mu \nu$ decays can serve as one probe of lepton universality
\cite{Hou,Hewett}.

\overfullrule 0pt As $|V_{ud}|$ has been quite accurately measured
in super-allowed $\beta$ decays \cite{Vud}, with a value of
0.97425(22) \cite{BM}, measurements of $\Gamma(\pi^+ \to \mu^+{\nu})$ yield a
value for $f_{\pi}$. Similarly, $|V_{us}|$ has been well measured in
semileptonic kaon decays, so a value for $f_{K}$ from $\Gamma(K^-
\to \mu^- \bar{\nu})$ can be compared to theoretical calculations.
Lattice gauge theory calculations, however,  have been
claimed to be very accurate in determining $f_K$, and these have
been used to predict $|V_{us}|$ \cite{Jutt}.

\section{Charmed mesons}
We review current measurements, starting with the charm system.  The CLEO
collaboration has performed the only measurement of the branching fraction for
$D^+\to\mu^+\nu$ \cite{fD}. CLEO uses $e^+e^-$ collisions at the $\psi(3770)$
resonant energy where $D^-D^+$ pairs are copiously produced. They fully
reconstruct one of the $D$'s, find a candidate muon track of opposite sign to
the tag, and then use kinematical constraints to infer the existence of a
missing neutrino and hence the $\mu\nu$ decay of the other $D$. They find
${\cal{B}}(D^+\to\mu^+\nu) = (3.82 \pm 0.32 \pm 0.09) \times 10^{-4}$.
We use the well-measured $D^+$ lifetime of 1.040(7) ps, and assuming $|V_{cd}|$
equals $|V_{us}|=0.2246(12)$ \cite{BM}  minus higher order correction terms
\cite{Charles}, we find $|V_{cd}|=0.2245(12)$.  The CLEO branching fraction
result then translates into a value of
$$
f_{D^+}=(206.7\pm 8.5\pm 2.5)~{\rm MeV}~.
$$
This result includes a 1\% correction (lowering) of the rate due to the
presence of the radiative $\mu^+\nu\gamma$ final state based on the estimate by
Dobrescu and Kronfeld \cite{Kron}.

Before we compare this result with theoretical predictions, we
discuss the $D_s^+$. Measurements of $f_{D_s^+}$ have been made by
several groups and are listed in Table~\ref{tab:fDs}
\cite{CLEO-c,Belle-munu,CLEO-rho,CLEO-CSP,Sanchez}.  We exclude older values
obtained by normalizing to $D_s^+$ decay modes
that are not
well defined. Many measurements, for example, used the $\phi\pi^+$ mode. This
decay is a subset of the $D_s^+\to K^+ K^- \pi^+$ channel which has
interferences from other modes populating the $K^+K^-$ mass region near the
$\phi$, the most prominent of which is the $f_0(980)$. Thus the extraction of
effective $\phi\pi^+$ rate is sensitive to the mass resolution of the
experiment and the cuts used to define the $\phi$ mass region
\cite{reason,Babar-munu}.
The CLEO, BaBar and Belle $\mu^+ \nu$ results rely on fully reconstructing all the final
state particles except for the neutrino and using a missing-mass technique to
infer the existence of the neutrino. CLEO uses $e^+e^-\to D_sD_s^*$ collisions
at 4170 MeV, while Babar and Belle use $e^+e^-\to D Kn\pi D_s^*$ collisions at
energies near the $\Upsilon(4S)$.

When selecting the $\tau^+\to\pi^+\bar{\nu}$ and $\tau^+\to\rho^+\bar{\nu}$
decay modes, CLEO uses both calculation of the missing-mass and the fact that
there should be no extra energy in the event beyond that deposited by the
measured tagged $D_s^-$ and the $\tau^+$ decay products. The $\tau^+\to
e^+\nu\bar{\nu}$ mode, however, uses only extra energy. 
BaBar measures $\Gamma(D_s^+ \to \tau^+ \nu)/\Gamma(D_s^+ \to \overline{K}^0
K^+)$ using the $\tau^+ \to e^+ \nu \bar \nu$ mode.

\begin{table}[htb]
\caption{Experimental results for ${\cal{B}}(D_s^+\to \mu^+\nu)$, ${\cal{B}}
(D_s^+\to \tau^+\nu)$, and $f_{D_s^+}$. Numbers for $f_{D_s^+}$ have been
extracted using updated values for masses and $|V_{cs}|$ (see text). Systematic
uncertainties for errors on the $D_s^+$ lifetime and mass are included;
radiative corrections have been included. Common systematic errors in the CLEO
results have been taken into account.
\label{tab:fDs}}
\begin{center}
\begin{tabular}{llccc}
 \hline\hline
& Experiment & Mode & ${\cal{B}}$ & $f_{D_s^+}$ (MeV)\\ \hline
& CLEO-c \cite{CLEO-c}& $\mu^+\nu$& $(5.65\pm
0.45\pm 0.17)\times 10^{-3}$ & $257.6\pm 10.3\pm 4.3$\\
& BaBar \cite{Sanchez}& $\mu^+\nu$& $(6.02\pm
0.38\pm 0.34)\times 10^{-3}$ & $265.9\pm 8.4\pm 7.7$\\
&Belle \cite{Belle-munu}
& $\mu^+\nu$ & $(6.38\pm 0.76\pm 0.57)\times 10^{-3}$ & $274\pm 16 \pm 12 $ \\
\hline
& Average & $\mu^+\nu$ & $(5.89\pm 0.33)\times 10^{-3}$ & $263.0\pm 7.3$ \\
\hline
& CLEO-c \cite{CLEO-c} & $\tau^+\nu~(\pi^+\overline{\nu})$ & $(6.42\pm 0.81 \pm
0.18) \times 10^{-2}$ & $278.0\pm 17.5 \pm 4.4 $ \\
& CLEO-c \cite{CLEO-rho} & $\tau^+\nu~(\rho^+\overline{\nu})$ & $(5.52\pm 0.57
\pm 0.21)\times 10^{-2}$ & $257.8\pm 13.3 \pm 5.2 $ \\
& CLEO-c \cite{CLEO-CSP} & $\tau^+\nu~(e^+\nu\overline{\nu})$ &
$(5.30\pm 0.47\pm 0.22)\times 10^{-2}$ & $252.6\pm 11.2 \pm 5.6 $ \\
& BaBar \cite{Sanchez} & $\tau^+\nu~(e^+(\mu^+)\nu\overline{\nu})$ &
$(5.00\pm 0.35\pm 0.49)\times 10^{-2}$ & $245.4\pm 8.6 \pm 12.2 $ \\ \hline
&Average & $\tau^+\nu$ & $(5.43\pm 0.31)\times 10^{-2}$ & $255.7\pm 7.2$
\\ \hline\hline
\end{tabular}
\end{center}
\end{table}

We extract the decay constant from the measured branching ratios using the
$D_s^+$ mass of 1.96847(33) GeV,  the $\tau^+$ mass of 1.77682(16) GeV, and a
lifetime of 0.500(7) ps.  We use the first order correction $|V_{cs}| =
|V_{ud}| - |V_{cb}|^2/2$ \cite{Charles} ; taking $|V_{ud}| = 0.97425(22)$
\cite{Vud}, and $|V_{cb}| =0.04$ from an average of exclusive and inclusive
semileptonic $B$  decay results as discussed in Ref.~\cite{Vcb}, we find
$|V_{cs}| = 0.97345(22)$.   CLEO has
included the radiative correction of 1\% in the $\mu^+\nu$ rate listed in the
Table \cite{Kron}~(the $\tau^+\nu$ rates need not be corrected). Other
theoretical calculations show that the $\gamma\mu^+\nu$ rate is a factor of
40--100 below the $\mu^+\nu$ rate for charm \cite{theories-rad}. As this is a
small effect we do not attempt to correct the other measurements.

The average decay constant cannot simply be obtained by averaging the values
in Table~\ref{tab:fDs} since there are correlated errors between the $\mu^+\nu$
and $\tau^+\nu$ values.  Table \ref{tab:both} gives the average values of
$f_{D_s}$ where the experiments have included the correlations.

\begin{table}[htb]
\caption{Experimental results for $f_{D_s^+}$ taking into account the common
systematic errors in the $\mu^+\nu$ and $\tau^+\nu$ measurements.}
\label{tab:both}
\begin{center}
\begin{tabular}{lc}
 \hline\hline
Experiment &$f_{D_s^+}$ (MeV)\\\hline
CLEO-c & $259.0\pm 6.2\pm 3.0$\\
BaBar & $258.8\pm 6.4\pm 7.5$\\
Belle & $273.8\pm 16.3\pm 12.2$\\\hline
Average of $\mu^+\nu+\tau^+\nu$ & $260.0\pm 5.4$
\\ \hline\hline
\end{tabular}
\end{center}
\end{table}

Our experimental average is
$$
f_{D_s^+}=(260.0\pm 5.4){\rm ~MeV}.
$$
Furthermore, the ratio of branching fractions is found to be 
\begin{equation}
R\equiv \frac{{\cal{B}}(D_s^+\to \tau^+\nu)}{{\cal{B}}(D_s^+\to \mu^+\nu)} =
9.2\pm 0.7,
\end{equation}
where a value of 9.76 is predicted in the Standard Model. Assuming lepton
universality then we can derive improved values for the leptonic decay
branching fractions of
\begin{eqnarray}
{\cal{B}}(D_s^+\to \mu^+\nu)&=&(5.75\pm0.24)\times 10^{-3},~~{\rm and}\nonumber\\
{\cal{B}}(D_s^+\to \tau^+\nu)&=&(5.61\pm0.24)\times 10^{-2}~.
\end{eqnarray}

The experimentally determined ratio of decay constants is
$f_{D_s^+}/f_{D^+}=1.26\pm 0.06$.

\begin{table}[htb]
\caption{Theoretical predictions of $f_{D^+_s}$, $f_{D^+}$, and
$f_{D_s^+}/f_{D^+}$.  Quenched lattice calculations are omitted, while
PQL indicates a partially-quenched lattice calculation.
(Only selected results having errors are included.)
\label{tab:Models}}
\begin{center}
\begin{tabular}{llccc}
\hline\hline
& Model & $f_{D_s^+}$(MeV) & $f_{D^+}$(MeV) & $f_{D_s^+}/f_{D^+}$\\\hline
& Experiment (our averages)& $260.0 \pm 5.4$ &
$206.7\pm 8.9$& $1.26\pm 0.06$ \\ \hline
 & Lattice (HPQCD) \cite{Lat:Dav} & $248.0\pm2.5$ & $213\pm4$ & $1.164\pm
 0.018$ \\
 & Lattice (FNAL+MILC) \cite{Lat:Milc} &
 $260.1\pm10.8$ & $218.9\pm11.3$ & $1.188\pm0.025$ \\
&PQL \cite{Lat:Nf2}& $244\pm 8$&$197\pm 9$&
   $1.24\pm 0.03$\\
& QCD sum rules \cite{Bordes} & $205\pm 22$& $177\pm 21$& $1.16\pm 0.01\pm0.03$
\\
& QCD sum rules \cite{Lucha} & $245.3\pm15.7\pm4.5$ & $206.2\pm7.3\pm5.1$ &
 $1.193\pm0.025\pm0.007$ \\
& Field correlators \cite{Field} & $260\pm 10$& $210\pm 10$& $1.24\pm 0.03$\\
& Light front \cite{LF} & $268.3\pm 19.1$ & 206 (fixed) & $1.30\pm 0.04$\\
\hline\hline
\end{tabular}
\end{center}
\end{table}

Table~\ref{tab:Models} compares the experimental $f_{D_s^+}$ with theoretical
calculations \cite{Lat:Dav,Lat:Milc,Lat:Nf2,Bordes,Lucha,Field,LF}.
While most theories give values lower than the $f_{D_s^+}$ measurement, the
errors are sufficiently large, in most cases, to declare success.  The largest
discrepancy (2.0 standard deviations) is with an unquenched lattice calculation
\cite{Lat:Dav}.

Upper limits on $f_{D^+}$  and $f_{D_s}$ of 230 and 270 MeV, respectively, have
been determined using two-point correlation functions by Khodjamirian
\cite{Kho}.  The $D^+$ result is safely below this limit, while the average
$D_s$ result is also, but older results \cite{Previous} not used in our average
are often above the limit.

Akeroyd and Chen \cite{AkeroydC} pointed out that leptonic decay widths are
modified in two-Higgs-doublet models (2HDM).  Specifically, for the $D^+$ and
$D^+_s$, Eq.~(\ref{equ_rate}) is modified by a factor $r_q$ multiplying the
right-hand side \cite{AkeroydM}:


$$
r_q=\left[1+\left(1\over{m_c+m_q}\right)\left({M_{D_q}\over M_{H^+}}\right)^2
\left(m_c-\frac{m_q\tan^2\beta}{1+\epsilon_0\tan\beta}\right)\right]^2,
$$

\noindent where $m_{H^+}$ is the charged Higgs mass, $M_{D_q}$ is
the mass of the $D$ meson (containing the light quark $q$), $m_c$ is
the charm quark mass, $m_q$ is the light-quark mass, and $\tan\beta$
is the ratio of the vacuum expectation values of the two Higgs
doublets. In models where the fermion mass arises from coupling to more
than one vacuum expectation value $\epsilon_0$ can be non-zero, perhaps
as large as 0.01. For the $D^+$, $m_d
\ll m_c$, and the change due to the $H^+$ is very small. For the
$D_s^+$, however, the effect can be substantial.

A major concern is the need for the Standard Model (SM) value of $f_{D_s^+}$.
We can take that from a theoretical model. Our most aggressive choice is that
of the unquenched lattice calculation \cite{Lat:Dav}, because it claims the
smallest error.  Since the charged Higgs would lower the rate compared to the
SM, in principle, experiment gives a lower limit on the charged Higgs mass.
However, the value for the predicted decay constant using this model is 2.0
standard deviations {\it below} the measurement.  If this small discrepancy
is to be taken seriously, either (a) the
model of Ref.~\cite{Lat:Dav} is not representative; (b) no value of $m_{H^+}$
in the two-Higgs doublet model will satisfy the constraint at 99\% confidence
level; or (c) there is new physics, different from the 2HDM, that interferes
constructively with the SM amplitude such as in the R-parity-violating model of
Akeroyd and Recksiegel \cite{Rviolating}.

To sum up, the situation is not clear. To set limits on new physics we need an
independent calculation of $f_{Ds}$ with comparable accuracy, and more precise
measurements would also be useful.

\section{\boldmath\bf The $B$ meson}
The Belle and BaBar collaborations have found evidence for $B^-\to\tau^-
\overline{\nu}$ decay in $e^+e^-\to B^-B^+$ collisions at the $\Upsilon(4S)$ energy.
The analysis relies on reconstructing a hadronic or semi-leptonic $B$ decay
tag, finding a $\tau$ candidate in the remaining track and or photon
candidates,  and examining the extra energy in the event which should be close
to zero for a real $\tau^-$ decay to $e^- \nu \bar \nu$ or $\mu^- \nu \bar \nu$
opposite a $B^+$ tag. The results are listed in Table~\ref{tab:Btotaunu}.

\begin{table}[htb]
\caption{Experimental results for ${\cal{B}}(B^-\to \tau^-\overline{\nu})$.
We have computed an average for the two Belle measurements assuming that the
systematic errors are uncorrelated.\label{tab:Btotaunu}}
\begin{center}
\begin{tabular}{lllc} \hline\hline
&Experiment & Tag &${\cal{B}}$ (units of $10^{-4}$)\hfill\\
\hline
&Belle~\cite{BelleH}&Hadronic&$1.79\,^{+0.56\,+0.46}_{-0.49\,-0.51}$\\
&Belle~\cite{BelleS}&Semileptonic&$1.54\,^{+0.38\,+0.29}_{-0.37\,-0.31}$\\
&Belle&Our average&$1.62 \pm 0.40$ \\
&BaBar~\cite{BaBarH} & Hadronic & $1.80\,^{+0.57}_{-0.54}\pm0.26$\\
&BaBar~\cite{BaBarS} & Semileptonic & $1.7\pm 0.8\pm 0.2$\\
&BaBar & Average \cite{BaBarH} & $1.76 \pm 0.49$\\\hline
& &Our average & $1.68\pm0.31$\\
\hline\hline
\end{tabular}
\end{center}
\end{table}

There are large backgrounds under the signals in all cases. The systematic
errors are also quite large, on the order of 20\%. Thus, the significances are
not that large.  Belle quotes 3.5$\sigma$ and 3.6$\sigma$ for their hadronic
and semileptonic tags, respectively, while BaBar quotes 3.3$\sigma$ and 2.3%
$\sigma$, again for hadronic and semileptonic tags. We note that the four
central values are remarkably close to the average considering the large
errors on all the measurements.
More accuracy would be useful to investigate the effects of new physics.

\begin{figure}[hbt]
\centering
\includegraphics[width=80mm]{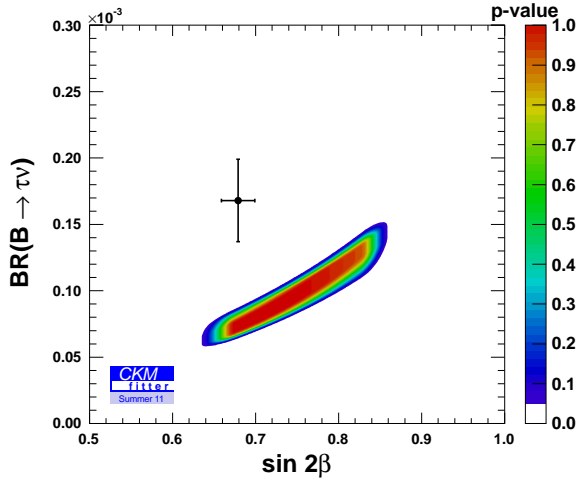}
\vspace{-4mm}
\caption{Measured versus predicted values of ${\cal{B}}(B^-\to\tau^-
\overline{\nu})$ versus $\sin2\beta$ from the CKM fitter group. The point with
error bars shows the measured values, while the predictions are in colors, with
the color being related to the confidence level. (From the CKM Fitter group.)}
\label{sin2b_Btaunu}
\end{figure}

We extract a SM value using Eq.~(\ref{equ_rate}). Here theory provides a value
of $f_B=(194\pm 9)$ MeV \cite{fBl}.
We also need a value for $|V_{ub}|$. Here significant differences
arise between using inclusive charmless semileptonic decays and
the exclusive decay $B\to\pi\ell^+\nu$ \cite{ABS}.  The inclusive decays give
rise to a value of $|V_{ub}|=(4.27\pm 0.38)\times 10^{-3}$  while the exclusive
measurements yield $|V_{ub}|=(3.38 \pm 0.36)\times 10^{-3}$, where the errors
are dominantly theoretical \cite{KM}.  Their average,
enlarging the error in the standard manner because the results differ, is
$|V_{ub}|=(3.80\pm0.44)\times 10^{-3}$. Using these values and the PDG values
for the $B^+$ mass and lifetime, we arrive at the SM prediction for the
$\tau^-\bar{\nu}$ branching fraction of $(0.96\pm 0.24)\times 10^{-4}$. This
value is about a factor of two smaller than the measurements.  There is a 6.6\%
probability that the data and the SM prediction are consistent. This difference
is more clearly seen by examining the correlation between the CKM angle $\beta$
and ${\cal{B}}(B^-\to\tau^-\bar{\nu})$. The CKM fitter group provides a fit to a
large number of measurements involving heavy quark transitions
\cite{CKMfitter}.  The point in Fig.~\ref{sin2b_Btaunu} shows the directly
measured values, while the predictions from their fit without the direct
measurements are also shown.  There is about a factor of two discrepancy
between the measured value of ${\cal{B}}(B^-\to\tau^-\overline{\nu})$ and the
fit prediction.

\section{Charged pions and kaons}
We now discuss the determination of charged pion and kaon decay constants.
The sum of branching fractions for
$\pi^- \to \mu^- \bar \nu$ and $\pi^- \to \mu^- \bar \nu \gamma$ is
99.98770(4)\%.  The two modes are difficult to separate experimentally, so we
use this sum, with Eq.~(\ref{equ_rate}) modified to include photon emission
and radiative corrections \cite{Marciano-Sirlin}.  The branching fraction
together with the lifetime 26.033(5) ns gives
$$
 f_{\pi^-} = (130.41\pm 0.03\pm 0.20)~{\rm MeV}~.
$$
\noindent The first error is due to the error on $|V_{ud}|$,
0.97425(22) \cite{Vud}; the second is due to the higher-order
corrections, and is much larger.

Similarly, the sum of branching fractions for $K^- \to \mu^- \bar
\nu$ and $K^- \to \mu^- \bar \nu \gamma$ is 63.55(11)\%, and the
lifetime is 12.3840(193) ns \cite{Flavi}.  Measurements of semileptonic kaon
decays provide a value for the product $f_+(0)|V_{us}|$, where $f_+(0)$ is the
form-factor at zero four-momentum transfer between the initial state kaon and
the final state pion. We use a value for $f_+(0)|V_{us}|$ of 0.21664(48)
\cite{Flavi}. The $f_+(0)$ must be determined theoretically. We follow Blucher
and Marciano \cite{BM} in using the lattice calculation $f_+(0)=0.9644\pm
0.0049$ \cite{latticefp}, since it appears to be more precise than the classic
Leutwyler-Roos calculation $f_+(0)=0.961\pm 0.008$ \cite{LR}.
[Other recent averages are $0.956 \pm 0.008$ \cite{Colangelo} and $0.9588
\pm 0.0044$ \cite{latticeav}.]  Using the value from Ref.\ \cite{latticefp},
the result is $|V_{us}|=0.2246\pm 0.0012$, consistent with the hyperon decay
value of $0.2250\pm 0.0027$ \cite{hyperon}.  We derive
$$
f_{K^-} =(156.1 \pm 0.2\pm 0.8\pm 0.2)~{\rm MeV}~.
$$

\noindent The first error is due to the error on $\Gamma$; the second is due to
the CKM factor $|V_{us}|$, and the third is due to the higher-order
corrections. The largest source of error in these corrections
depends on the QCD part, which is based on one calculation in the
large $N_c$ framework.  We have doubled the quoted error here; this
would probably be unnecessary if other calculations were to come to
similar conclusions.  A large part of the additional uncertainty
vanishes in the ratio of the $K^-$ and $\pi^-$ decay constants, which is
$$
f_{K^-}/f_{\pi^-} = 1.197 \pm 0.002 \pm 0.006 \pm 0.001~.
$$

\noindent
The first error is due to the measured decay rates; the second is due to the
uncertainties on the CKM factors; the third is due to the uncertainties in the
radiative correction ratio.

These measurements have been used in conjunction with calculations of
$f_K/f_{\pi}$ in order to find a value for $|V_{us}|/|V_{ud}|$.  Three recent
lattice predictions of $f_K/f_{\pi}$ are
$1.189 \pm 0.007$ \cite{Lat:Foll},
$1.192 \pm 0.007 \pm 0.006$ \cite{fkfpi}, and
$1.197 \pm0.002^{+0.003}_{-0.007}$ \cite{Bazavov},
yielding an average by the FLAG group of $1.195 \pm 0.005$ \cite{Colangelo}.
[There is also a new value $1.1872 \pm 0.0041$ (statistical errors only)
\cite{MILClite}].  Together with the precisely measured $|V_{ud}|$, this gives
an independent measure of $|V_{us}|$ \cite{Jutt,Flavi}.

We gratefully acknowledge support of the U. S. National Science Foundation
and the U. S. Department of Energy through Grant No.\ DE-FG02-90ER40560.
We thank C. Davies, A. Kronfeld, W. Marciano, S. Sharpe, R. Van de Water,
and H. Wittig for helpful advice.

\end{document}